# Title: Nanostructured intense-laser cleaner

**Authors:** X. F. Li[1,2*], S. Kawata[2], Q. Kong[1], P. X. Wang[1], Q. Yu[1,3], Y. J. Gu[1,3] and J. F. Qu[1]

**Affiliations:**

[1]Institute of Modern Physics, Fudan University, Shanghai 200433, People's Republic of China.

[2]Graduate School of Engineering, Utsunomiya University, 7-1-2 Yohtoh, Utsunomiya 321-8585, Japan.

[3]Institute of Physics of the ASCR, ELI-Beamlines Project, Na Slovance 2, 18221 Prague, Czech Republic.

Corresponding author: xiaofengli13@fudan.edu.cn

**Abstract:** A nanostructured target is proposed to enhance an intense-laser contrast: when a laser prepulse irradiates a nanostructured solid target surface, the prepulse is absorbed effectively by the nanostructured surface. The nanostructure size should be less than the laser wavelength. After the prepulse absorption, the front part of the main pulse destroys the microstructure and makes the surface a flat plasma mirror. The body of the main pulse is reflected almost perfectly. Compared with conventional plasma mirrors, the nanostructured surface is effective for the absorption of the intense laser prepulse, whose intensity is higher than $10^{14}$ W/cm$^2$. The nanostructured laser cleaner improves the laser pulse contrast by about a hundredfold. The nanostructured laser cleaner works well for next-generation intense lasers.

**One Sentence Summary:** A remarkable laser-cleaning device was discovered by computer simulations, and it reduces an undesirable prepulse significantly to produce clean ultra-intense lasers.

## I. Introduction

The laser intensity has rapidly been increasing in these days especially by the chirped pulse amplification (CPA)(*1*) technique. By the intense and short pulse lasers, high-energy mono-energetic electron beams are obtained, for example, by the laser wakefield acceleration (*2-6*). The laser ion acceleration(*7-12*) and high-energy-density plasma physics(*13, 14*) have also been developed. As the laser peak intensity attainable increases, the requirement for the laser contrast ratio between the prepulse and the main pulse becomes severe for experiments and scientific applications. A number of experiments on intense-laser matter interaction require a sharp polished solid density at the arrival of the main laser pulse, and it is particularly true for ultrathin foil targets used in laser particle-acceleration schemes. The preceding intense prepulse may produce an undesired pre-plasma, which would interfere with the study on the laser-matter interaction.

The laser intensity contrast is defined as the prepulse intensity divided by that of the main pulse, and is used to characterize the intensity cleanness of the laser pulses(*15*). In order to improve the laser contrast, there are several methods proposed: second harmonic generation after a laser pulse compression(*16*), polarization wave generation(*17*), relativistic plasma shutters(*18*) and plasma mirror(*14, 19, 20*). By the methods proposed, the laser pulse contrast has been improved to about $10^{-6}$ to $10^{-8}$. In order to obtain the higher contrast ratio, the plasma mirror(*14*) improves the pulse contrast to $10^{-8}$. By the double plasma mirrors(*19, 21*), the pulse contrast ratio was improved to $10^{-10}$. However, when the laser intensity reaches more than $10^{21} \sim 10^{24} \text{W/cm}^2$, the prepulse intensity would be $10^{13} \sim 10^{16} \text{W/cm}^2$, which is more than two orders of the magnitude higher than the ionization threshold of any target materials.

The plasma mirror may be an ultrafast shutter switched by the laser light itself (*15, 19-23*). The prepulse penetrates the material of the plasma mirror, though the main pulse makes the mirror surface a flat plasma and is reflected by the plasma layer at the plasma mirror surface. The plasma mirror design depends on the laser intensity. However, the prepulse itself makes the plasma mirror surface a plasma, when the laser

intensity(*13*) reaches about $10^{24}$ W/cm$^2$ or higher with the prepulse intensity larger than $10^{16}$ W/cm$^2$. The plasma mirrors do not work, when the laser intensity becomes very high in the near future. In this paper, a solid target with a nanostructured surface is employed to improve the laser intensity contrast for the next-generation intense lasers. The prepulse is absorbed by the nanostructured surface, and the major part of the main pulse is reflected by the target surface plasma, which is created by the head of the main laser pulse.

**II. Laser cleaning method**

In this paper, with the incidence angle of $\theta = 45^0$ a laser prepulse and a successive main pulse irradiate a solid target, which has a nanostructured surface, as shown in Fig. 1(a). Firstly the laser prepulse interacts with the nanostructured surface. The nanostructured surface consists of many quadrilateral wings in this work as an example nanostructure. It should be noted that the wing shape is not essential, though the fine structure size should be small compared with the laser wavelength to absorb the prepulse effectively (*9-11*). Here, the nanostructure wing width is specified by *δ*, and the distance between the two nanostructure wings is Δ, and the nanostructure length is *L* (see Fig. 1(a)). As shown in Fig. 1(b), the most part of the prepulse is scattered among the fine wings several times, and interacts with the nanostructured surface well. Then the prepulse is absorbed well by the nanostructured surface. During the interaction, the nanostructure does not change and keeps its structure until the main pulse head reaches the surface. When the main pulse arrives at the nanostructured surface as presented in Fig. 1(b), the small front part of the main pulse destroys the nanostructure, and the target surface becomes a smooth plasma surface as shown in Fig. 1(c) schematically. Then the body part of the main pulse is reflected by the smooth plasma surface. By the mechanism proposed in this paper, the laser intensity contrast is improved significantly for the next-generation ultra-intense lasers.

We study the nanostructured intense-laser cleaner by 2.5D particle-in-cell simulations(*24*). The laser pulse is linearly polarized in the *y* direction with the Gaussian focused profile:

$$E = E_0 \frac{w(x)}{w_0} \exp(-\frac{y^2+z^2}{w^2(x)}) \exp(-\frac{(kx-\omega_L t)^2}{(0.5\tau)^2}) \cos(\varphi), \quad (1)$$

where $w(x) = w_0 \left[1+(x-x_0)^2/z_R^2\right]^{0.5}$, $w_0$ is the laser waist size, $x_0$ the position of the laser waist, $z_R = \pi w_0^2/\lambda$ the Rayleigh length, and $\tau = 30 fs$ the pulse duration. The laser wavelength is $\lambda$=800nm, and the simulation resolution in space is $dx = dy = 0.02\lambda$. The laser beam is injected from the left boundary of the simulation box, and is focused on $x=20\lambda$. The laser intensity is $a_0 = eE_0/m_e\omega_L c$ and the laser waist is $w_0 = 5\lambda$, where $m_e$ is the electron mass and $\omega_L$ the laser frequency.

The Aluminum substrate is employed to reduce the target surface deformation. The Al solid density is $42n_c$, where $n_c$ is the critical density. The prepulse interacts with and are absorbed by the nanostructured surface installed on the Al surface. Then the Aluminum interaction surface is quickly ionized by the head of the main pulse. The last two electrons in the Al ground energy state are difficult to be ionized, and so the ionization degree employed in our simulations is 11. The material of the nanostructure should be selected depending on the laser intensity. In this work, Hydrogen is employed as an example nanostructured material, which represents a Hydrogen-rich plastic.

The laser pulse irradiates the nanostructured target in the x direction with the incidence angle of $\theta = 45^0$, and is reflected in the y direction. The laser reflection is measured by $I/I_0$, where $I$ is the reflected laser intensity and $I_0$ the incident laser intensity. The prepulse laser has the same laser waist and the pulse duration as the main pulse(25).

### III. Laser cleaning results

When an intense laser (the main pulse intensity is $I_0 = 2.2\times 10^{22} W/cm^2$ and the laser contrast ratio is $10^{-6}$) interacts with the nanostructured target, the prepulse ($2.2 \times 10^{16} W/cm^2$) is first absorbed well by the nanostructure, and the maximum intensity of the reflected prepulse is ~1% of the prepulse initial intensity, as presented in Fig. 2(a). However, when the main pulse is reflected at the nanostructured target surface, the main pulse keeps the same intensity as the original one, as shown in Fig. 2(b). In this case, the Hydrogen density $n_0$ is $n_0=30n_c$. The distance between the two nanostructure wings is $\Delta=0.9\lambda$ and its length is $L=2.5\lambda$. The width of the nanostructure wings is $\delta=0.1\lambda$. The laser contrast was improved from $10^{-6}$ to $10^{-8}$ by about two orders of magnitude, as presented in Fig. 2(c). The nanostructured target improves the laser contrast of the ultra-intense laser, at which the conventional plasma mirror does not work, with the same quality as one plasma mirror does.

In order to obtain the higher contrast ratio, it is necessary to design the appropriate nanostructured surface to absorb the prepulse energy effectively. Firstly, we used a low intensity laser interacting with the nanostructured target: the laser intensity is $a_0=0.10$, which corresponds to the prepulse intensity in Figs. 2. The front part of the laser reached on the target at 50fs, as shown in Fig. 3(a), and it leaves from the target at about 170fs. The maximal intensity of the reflected laser is ~1% of the initial intensity, as shown in Fig. 3(b). The major part of the prepulse laser was absorbed by the nanostructure. The distributions of the electron density and the proton density did not change remarkably, as presented in the Figs. 3(c) and (d), respectively. While the laser pulse was absorbed by the nanostructured surface, it did not change the nanostructure notably. The center of the small reflected pulse in this case is on $x=17.5\lambda$, as shown by the orange dotted line in Fig. 3(b). However, the previous result in Fig. 2 for the intense main laser pulse ($a_0=100$) shows that the center of the intense main pulse reflected is on $x=20.0\lambda$; the results in Figs. 2 and 3(b) demonstrate that the intense main pulse is reflected at the Al surface, though the prepulse is reflected at the nanostructure layer. As presented in Figs. 3(b) and (e), the laser intensity at $x=20.0\lambda$ ($I/I_0$~1.8%) is smaller than that at $x=17.5\lambda$ ($I/I_0$~2.3%). After the laser pulse interacts with the target, the prepulse location and that of the intense main pulse are different with each other in the nanostructured target, and it would be also benefit for the laser contrast improvement.

The quality of the reflected laser depends on the surface nanostructure. When the distance $\Delta$ of the nanostructure wing is same as the width $\delta$ of the wing ($\delta=\Delta=0.1\lambda$), 5%~18% of the laser intensity was reflected, though the reflection ratio is improved significantly to ~1% at $\delta=0.1\lambda$, $\Delta=0.9\lambda$ and $L=2.5\lambda$. With the increase in $\Delta$ ($\Delta=0.1\lambda$~$2.4\lambda$), the reflection intensity decreases firstly and then increases, as presented in Fig. 3(f). For a certain length ( $L$ ) of the nanostructure, there is the optimal value of $\Delta$ for the nanostructured surface to absorb the laser effectively. With the increase in the length ( $L$ ) of the nanostructure ($L=0.5\lambda$, $1.5\lambda$ and $2.5\lambda$), the laser energy was absorbed more. It could be explained as follows: when the distance $\Delta$ of the nanostructured wings is too narrow, the major part of the prepulse laser is reflected on the nanostructured surface, similar to a flat surface. On the other hand, when the distance $\Delta$ is too large, the laser reaches the Al surface and is reflected on the Aluminum solid target surface. When the length $L$ of the nanostructure wings becomes longer, the electromagnetic wave of the laser is scattered and absorbed among the long wing spaces of the nanostructure.

When the laser intensity is relatively higher ($a_0=100.0$, which corresponds to the main pulse intensity of ~$10^{22}$ W/cm$^2$ ), the most part of the laser pulse is reflected as shown in Fig. 4(a). The laser profile does not change significantly, and the maximum of the laser intensity is almost same with the incident laser initial intensity. On the laser axis, the maximum value of the reflected laser intensity also does not change, as shown in Fig.4 (b). It is demonstrated that the laser pulse keeps its original profile after the interaction with the nanostructured target. At the beginning of the interaction process ( at $t=90fs$ ), the nanostructured surface disappeared as shown in Figs. 4(c) and (d). The results in Fig. 4 show that the nanostructured surface is destroyed by the small front part of the intense laser; after that, the plasma surface reflects the main part of the intense laser completely.

When the laser intensity is relative low, the nanostructured surface absorbs the entire laser energy. However, the structure is destroyed by the head of the intense main laser. Therefore, we should choose the appropriate density of the nanostructure, which absorbs the prepulse effectively and reflects the major part of main pulse well. Figure 4(e) shows that the reflection ratio of the laser intensity as a function of the incident laser intensity with the different density of the nanostructured surface. We found that the lower intensity prepulse energy was absorbed well, and the lowest prepulse reflection ratio is 0.8%. However, when the main pulse laser intensity becomes higher, the most part of the laser intensity was reflected. With the increase in the nanostructure density, the more intense prepulse is absorbed effectively, as shown by a black dashed line ( $I/I_0$=3% ) in the inset of Fig. 4(e). For the soft density $n_0 = 10n_c$ and for the intensity $a_0 \sim 0.2$, the reflection ratio reaches 3%. For $n_0 = 40n_c$ and for the prepulse intensity $a_0 \sim 4$, the reflection ratio is ~3%.

We should use a soft material ($n_0 \leq 10n_c$) for the nanostructured surface, when the intensity of main pulse $a_0 < 20$. When the density of the nanostructured surface is $n_0 = 40n_c$ , it absorbs the prepulse $I_0 < 10^{19}$ W/cm$^2$ effectively. The nanostructured laser cleaner improves the intense laser contrast significantly, when the main pulse intensity is larger than ~10$^{20}$ W/cm$^2$. So far, when the laser main pulse intensity reaches 10$^{21}$ W/cm$^2$ or more, the double plasma mirrors have been used to improve the intensity contrast. When the laser intensity becomes larger than 10$^{22}$~10$^{23}$ W/cm$^2$, the prepulse would make the plasma mirror surface a plasma. The plasma mirror does not work in this case, though the nanostructured target proposed works well even for the super intense lasers.

In order to compare the intense main-laser reflection performance of the nanostructured target with that of the plasma mirror (a smooth solid surface without the nanostructure), one simulation result is presented in Figs. 4(b) and (f). The temporal profiles and the laser intensities of the reflected laser pulses are not different between the two cases. It is concluded that the nanostructured surface does not affect the main laser pulse profile significantly. The small front part of the main pulse was used to destroy the nanostructured surface, as shown by a green arrow in Fig. 4(b).

## IV. Conclusions

In this paper, the nanostructured laser cleaner was proposed, and the laser prepulse was well absorbed by the nanostructured target surface. It is found that the nanostructured laser target is quite effective for the next-generation super-intense laser pulse cleaning instead of the conventional plasma mirror. After the prepulse absorption, the main pulse head makes the nanostructured surface a flat plasma, and the body of the main laser pulse is reflected perfectly. The intense-laser contrast is remarkably improved especially for the higher intense prepulse of $\gg 10^{14}$ W/cm$^2$.

**Acknowledgements**

This work was partly supported by NSFC (No. 11175048), the Shanghai Nature Science Foundation (No. 11ZR1402700), and the Shanghai Scientific Research Innovation Key Projects (No. 12ZZ011). The work was also partly supported by the China Scholarship Council, Shanghai Leading Academic Discipline (Project B107), JSPS KAKENHI (Grant No. 15K05359), MEXT, JSPS, the ASHULA Project, ILE/Osaka University, CORE (Center for Optical Research and Education, Utsunomiya University, Japan), Fudan University, and CDI (Creative Dept. of Innovation, CCRD, Utsunomiya University)

**Figure Captions**

**Figure 1**

The schematic diagrams of the laser prepulse and main pulse interactions with the nanostructured target.

**Figure 2**

The profiles of the reflected laser prepulse (a) and the reflected main pulse (b) after the interactions with the nanostructured surface. (c) The laser pulse intensity contrast on the laser axis (a purple dashed line in (a) and (b)). The incident main laser intensity is $a_0 = 100$ and the intensity contrast ratio is $10^{-6}$. The laser pulse duration is $\tau = 30fs$ and the laser waist size is $w_0 = 5\lambda$. The prepulse has the same laser waist and pulse duration as the main pulse. The width ($\delta$), the distance ($\Delta$) and the length ($L$) of nanostructure is $0.1\lambda$, $0.9\lambda$ and $2.5\lambda$, respectively. The density of the nanostructured surface is $n_0 = 30n_c$.

**Figure3**

The low intensity laser interaction with the nanostructured surface. (a) The distribution of the incident laser intensity for the laser pulse at $50fs$ and $a_0 = 0.10$ (b) The profile of the reflected laser at $175fs$. The distribution of the electron density (c) and the proton density (d) for the nanostructured target at $175fs$. (e) The laser intensity in the different positions at $x = 17.5\lambda$ (a green line) and at $x = 20.0\lambda$ (a red line). (f) The reflection ratio ($I/I_0$) as a function of the nanostructure wing distance ($\Delta$) with the different length ($L$) of the nanostructured surface. The nanostructured surface density is $n_0 = 30n_c$.

**Figure 4**

The higher intensity laser interaction with the nanostructured surface. (a) The distribution of the reflected laser intensity at $175fs$. The initial intensity of the incident laser is $a_0 = 100.0$. (b) The laser intensity on $x = 20\lambda$ as a function of $y$ at $t = 175fs$ for the nanostructured surface (a blue line) and a solid target without the nanostructure (a red line). The distributions of the electron density (c) and the proton density (d) after the front part of laser pulse interaction with the nanostructured target at $t = 90fs$. (e)The reflection ratio as a function of the incident laser intensity with the different density nanostructured surface. (f) The reflected laser profile after the laser interacted with the solid target without the nanostructure. The laser parameter is same as in (a).

**Figure 1**

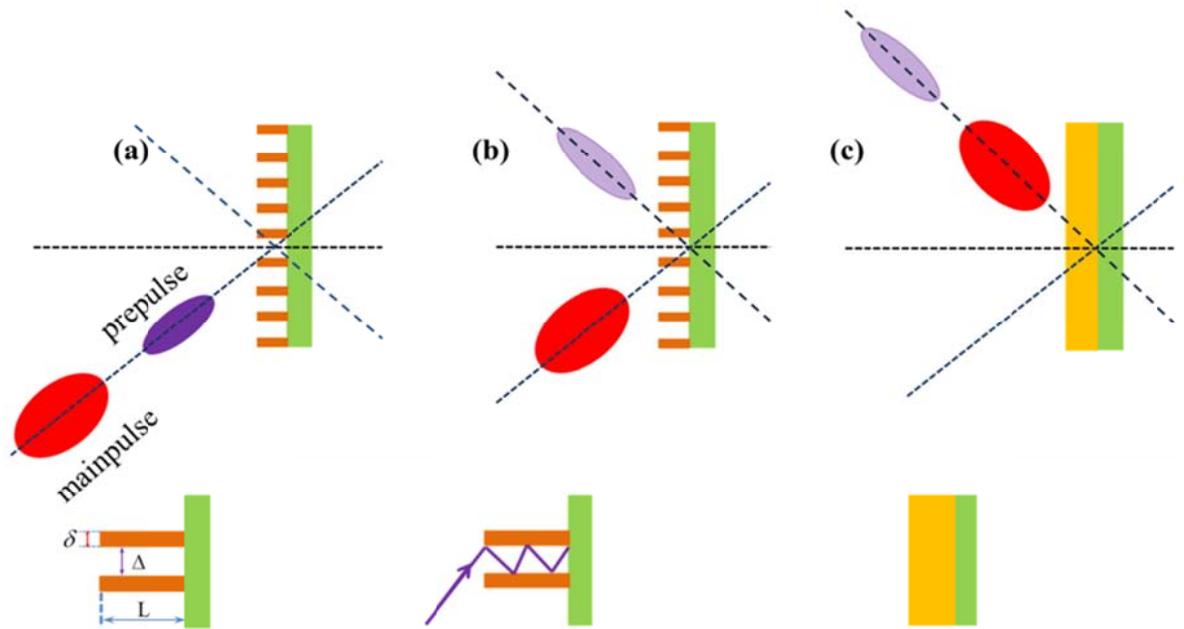

**Figure 2**

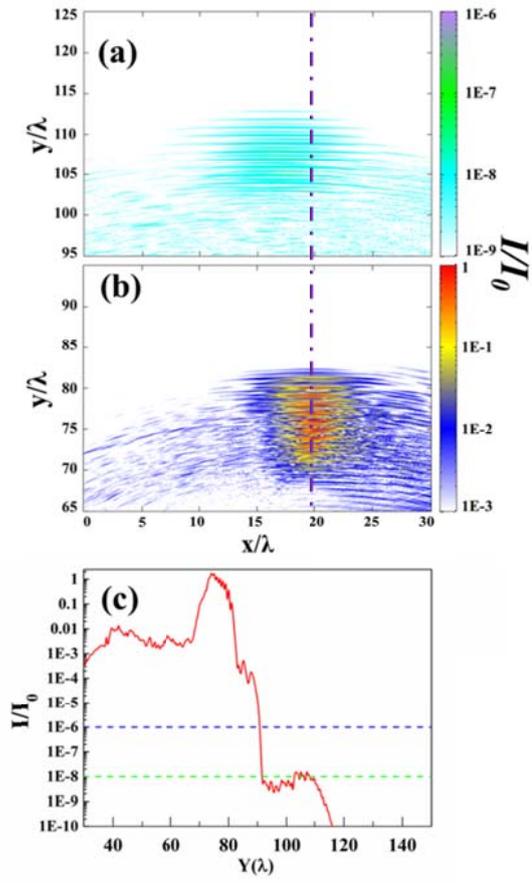

**Figure 3**

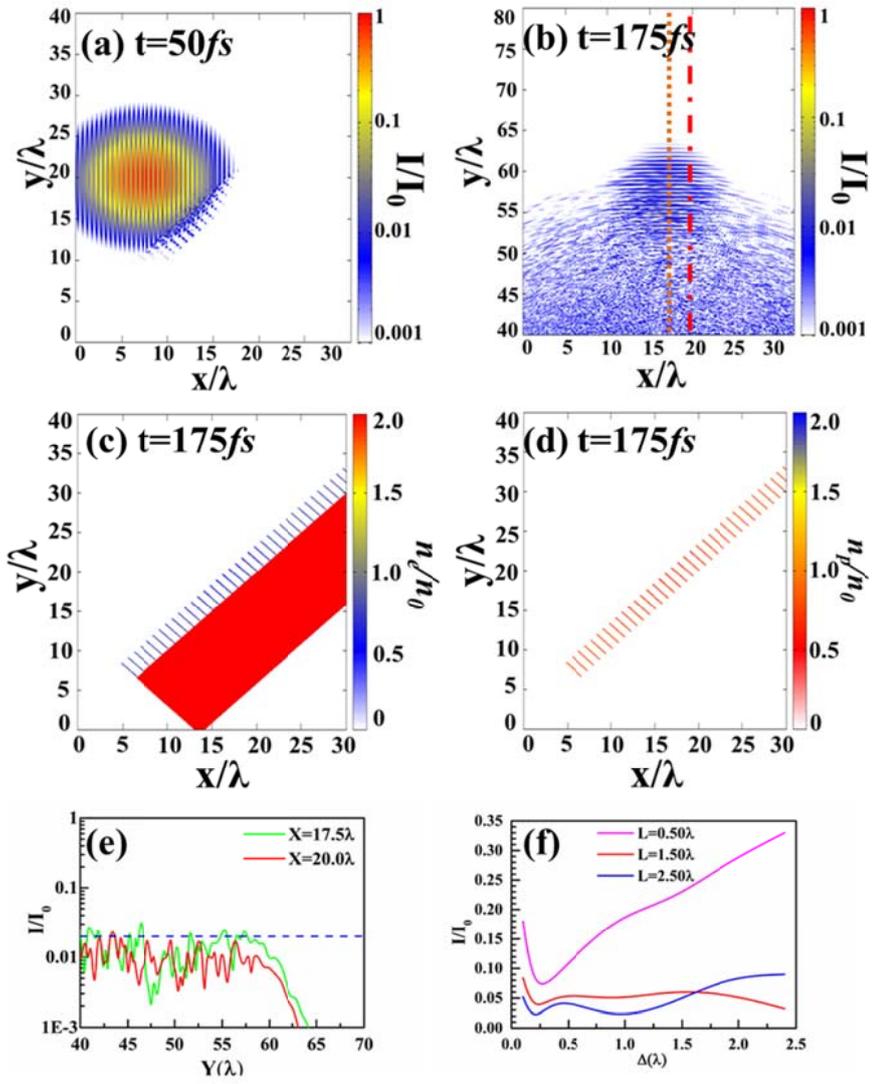

**Figure 4**

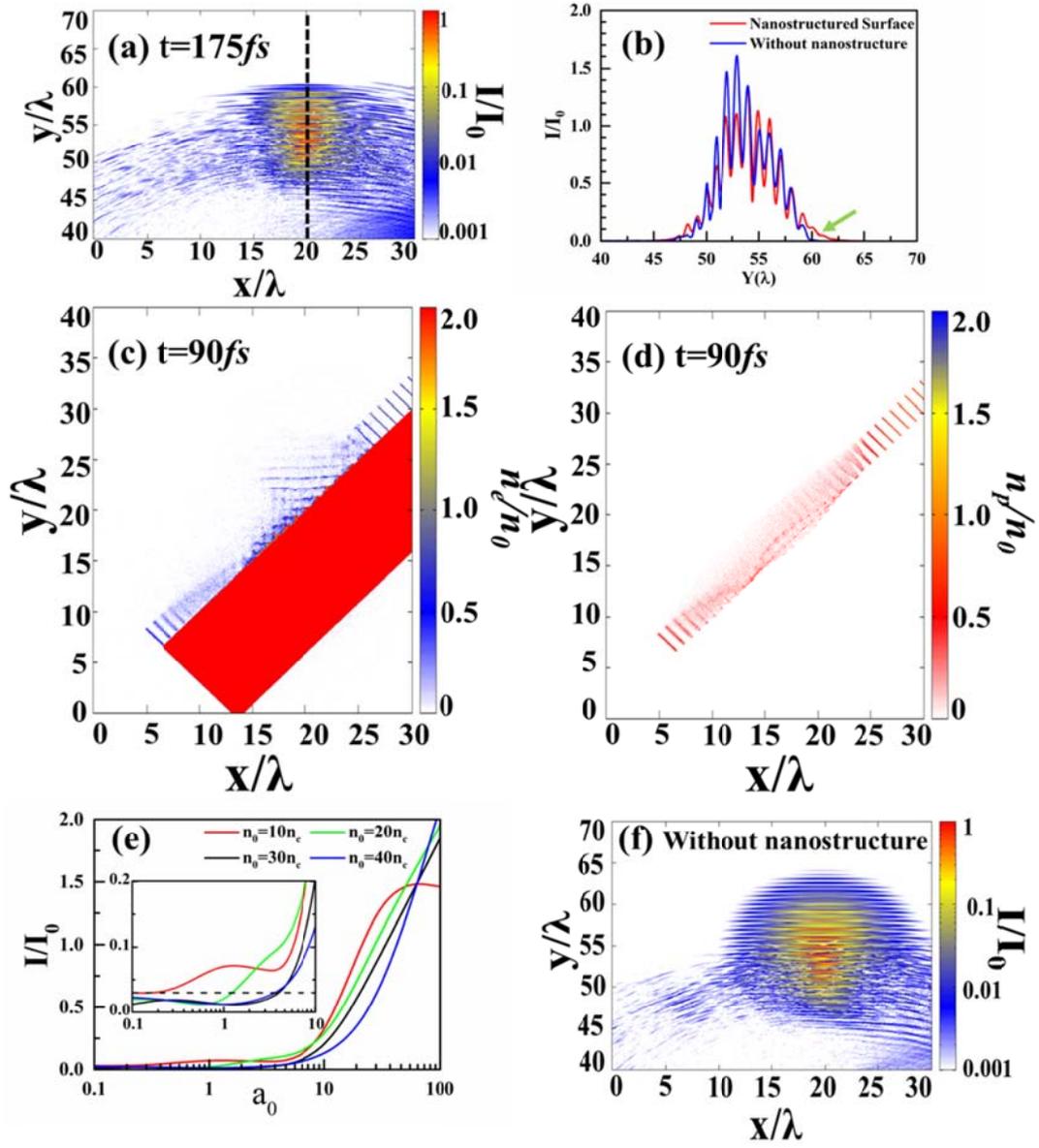